\documentclass[twocolumn]{aastex631}
\shorttitle{AASTeX v6.3.1 Sample article}
\shortauthors{Sie et al.}
\graphicspath{{./}{figures/}}

\begin{document}

\title{The key parameters controlling the photodesorption yield in interstellar CO ice analogs: Influence of ice deposition temperature and thickness}

\correspondingauthor{Y.-J. Chen}
\email{asperchen@phy.ncu.edu.tw}

\author[0000-0001-7755-7884]{N.-E., Sie}
\affil{Department of Physics, National Central University, Zhongli Dist., Taoyuan City 320317, Taiwan}

\author{Y.-T. Cho}
\affil{Department of Physics, National Central University, Zhongli Dist., Taoyuan City 320317, Taiwan}

\author[0000-0003-2741-2833]{C.-H. Huang}
\affiliation{Department of Physics, National Central University, Zhongli Dist., Taoyuan City 320317, Taiwan}

\author{G.M. Mu\~{n}oz Caro}
\affiliation{Centro de Astrobiolog\'ia (INTA-CSIC), Carretera de Ajalvir, km 4, Torrej\'on de Ardoz, 28850 Madrid, Spain}

\author{L.-C. Hsiao}
\affiliation{Department of Physics, National Central University, Zhongli Dist., Taoyuan City 320317, Taiwan}

\author{H.-C. Lin}
\affiliation{Department of Physics, National Central University, Zhongli Dist., Taoyuan City 320317, Taiwan}

\author[0000-0003-4497-3747]{Y.-J. Chen}
\affiliation{Department of Physics, National Central University, Zhongli Dist., Taoyuan City 320317, Taiwan}



\begin{abstract}
The overabundance of gas molecules in the coldest regions of space point to a non-thermal desorption process. Laboratory simulations show an efficient desorption of CO ice exposed to ultraviolet radiation, known as photodesorption, which decreases for increasing ice deposition temperature. However, the understanding of this abnormal phenomenon has remained elusive. In this work we show the same phenomenon, and in particular, a dramatic drop in the photodesorption yield is observed when the deposition temperature is 19 K and higher. Also the minimum ice thickness that accounts for a constant photodesorption yield of CO ice is deposition temperature dependent, an observation reported here for the first time. We propose that the key parameters that dominate the absorbed photon energy transfer in CO ice, and contribute to the measured photodesorption yields are the energy transfer length, single ice layer contributed desorption yield, and relative effective surface area. This set of parameters should be incorporated in astrophysical models that simulate photodesorption of the top CO-rich ice layer on icy dust populations with the size distribution which is ice thickness related.
\end{abstract}



\section{Introduction} \label{sec:intro}
Thermal desorption of molecules depleted in ice-covered dust grains is efficient in hot cores, but the low temperatures that reign in dark cloud interiors and radiation protected circumstellar regions invoke a different desorption mechanism to account for the observed molecular abundances in the gas phase. Along with other processes, such as cosmic ray bombardment and chemical desorption \citep{minissale2016dust, dartois2018cosmic}, photon-induced desorption or photodesorption is expected to contribute to the ejection of molecules from the ice in the cold regions of space. Although external ultraviolet (UV) photons do not permeate these environments, allowing dust temperatures to drop near 10 K and ice mantle accretion onto dust, the secondary UV photons produced by cosmic-rays driven hydrogen excitation can still impinge onto dust grains and drive photochemistry and photodesorption \citep{cecchi1992cosmic, shen2004cosmic}. In particular, carbon monoxide (CO) is one of the dominant gaseous species in the coldest regions and is also the dominant component of the top ice layer, with a second ice layer at the bottom dominated by water and other species \citep{pontoppidan2003m,pontoppidan2004mapping,boogert2015observations,penteado2015spectroscopic}. According to laboratory simulations, the photodesorption of CO ice is efficient and could contribute significantly to the CO gas abundances observed toward dense clouds \citep{G10}. The CO ice deposition temperature and deposition angle change the ice morphology and have an impact on the photodesorption yield \citep{oberg09, G16, Cristobal19}.

Nevertheless, the photodesorption yield of CO ice as a function of deposition temperature has not been sufficiently investigated. In \cite{oberg09}, the CO ice samples were deposited between 15 and 27 K, and UV irradiated at the temperature of deposition. The photodesorption yield of CO seems to decrease linearly with increasing deposition temperature, but the contribution of thermal desorption might lead to an overestimated photodesorption yield when irradiation proceeds at temperatures above 20 K. \cite{G16} shows that the photodesorption yield also decreases linearly when the deposition temperature is increased from 7 to 20 K (all irradiated at 7 K). We tried to use structure (, which is related to deposition temperature) to explain this phenomenon. We have observed the IR absorption band of CO ices shifting which is relative to the ice deposition temperature, and also succeed to explain this shifting is due to the spontelectric field ($E_S$) of CO ice, and furthermore shifting of vacuum-UV (VUV) CO ice absorption bands caused by Wannier-Mott exciton formation \citep{chen17}. However, we cannot explain the linear decrease in photodesorption yield with deposition temperature by ice structure or $E_S$ variations (VUV-absorption band shifting) since these two phenomena occur above 20 K. In the present work, the effect of deposition temperature of CO ice on photodesorption is explored from 12.5 K to 25 K in steps of 2 K to investigate in more detail this intriguing phenomenon. Furthermore, \cite{chen14} have reported that the different MDHL operation configurations, which produced different photon energy distributions, can cause the photodesorption yields of CO ice ranged from $(6.4\pm 0.2)\times 10^{-2}$ to $(2.1\pm 0.3)\times 10^{-1}$ molecules photon$^{-1}$. The incident photon intensity of different MDHL configurations are taken into account to calculate the wavelength weighted-absorption intensity, and the discrepancy of photodesorption yields of CO, $\rm CO_2$ and $\rm O_2$ between \cite{sie19} and \cite{martin15} can be explained after considering the absorption cross-section and the MDHL spectra of different configurations, which is a linear relation between the photodesorption yield and the integrated absorption intensity. 
\cite{G16} mentioned that the CO photodesorption rate values that have been reported by \cite{G10} and \cite{chen14}, respectively at 15 K and 14 K deposition temperatures, using a F-type microwave-discharged hydrogen UV lamp with comparable spectral emission, were $3.5\times 10^{-2}$ and $8.9\times 10^{-2}$ molecules per incident photon. If the factor 2.55 difference in their estimations of the total UV flux at the sample position, respectively $2.5\times 10^{14}$, measured using actinometry, and $9.8\times 10^{13}$ photons cm$^{-2}$ s$^{-1}$, measured with a calibrated Ni-mesh, is taken into account, both works converge exactly to the same value of the photodesorption rate.

The charge or energy transfer has been studied in previous literature, and the desorption induced by electronic transition (DIET) is proposed to dominate the photodesorption of CO ice, which is the process by which a molecule in the ice is photoexcited and the molecules near the surface can photodesorb \citep{reimann1988electronically,avouris1989fundamental,madey1994,bonn1999phonon,funk2000desorption,schwoerer2007organic,fayolle11, bertin2012uv}. So far we only know that the energy absorbed by molecules inside the ice can be transferred to the top few monolayers enabling photodesorption, but the parameters influencing the photodesorption yield were not sufficiently investigated.

In this work we propose three parameters involved in the photon energy transfer process in CO ice, namely, energy transfer length ($L(T)$), single ice layer contributed desorption yield ($Y(T)$), and relative effective surface area $R_A(T)$. Besides, the instantaneous photodesorption yields ($Y_{ipd}$) are calculated as a function of remaining CO ice column density (N$_{re}$(CO)) to figure out the contribution to photodesorption of these parameters, which are deposition temperature dependent.

\section{\label{sec:exp}Experimental set-up and procedures}
The experiments were conducted in the Interstellar Photoprocess System (IPS), which was described in detail in \cite{chen14}. In brief, IPS is an ultra-high vacuum (UHV) chamber with a base pressure of $\approx 3\times10^{-10}$ torr, consisting of a closed-cycle helium cryostat to cool down the CaF$_2$ substrate for ice deposition at 12.5--25 K in this work. The CO gas (purity 99.999 $\%$) is prepared in a pre-mixing system, flowing into the UHV chamber through a leak valve with pressure $4\times 10^{-9}$ torr during deposition. The column density of CO ice is monitored in situ by a transmission Fourier transform infrared spectrometer (FTIR), which is calculated by $N=\int_{band}^{} \frac{\tau_\nu d\nu}{A}$, where $N$ is the column density in molecules cm$^{-2}$, $\tau_\nu$ the optical depth of the band, $\nu$ the wavenumber in cm$^{-1}$, and $A$ is the band strength $1.1\times10^{-17}$ cm molecule$^{-1}$ adopted from \cite{jiang1975}. Each IR spectrum is taken every 3 minutes with a resolution of 2 cm$^{-1}$. A quadrupole mass spectrometer (QMS) is used to measure the desorbing molecules in the gas phase during irradiation. The band strength of CO ice deposited up to 25 K is constant, this was confirmed by another experiment where CO ice was deposited at 12.5 K and annealed to 25 K. In this experiment, the integrated absorbance did not change and lack of m/z=28 signal in the QMS indicates that desorption did not occur during annealing. The energetic VUV light source is generated by a microwave-discharge hydrogen-flow Lamp (MDHL) equipped with a MgF$_2$ window for an emission spectrum from 114--170 nm, and the photon flux is about $4.8\times 10^{13}$ photons cm$^{-2}$ s$^{-1}$, monitored in-situ by a calibrated Ni-mesh during irradiation periods. The UV emission spectrum generated by the MDHL has a great similarity to the secondary-UV spectrum in the interior of dense molecular clouds calculated in \cite{gredel1989}, providing a method to mimic the light source that processes ice mantles \citep{chen14,cruz14p}. The VUV irradiation processes are all performed at 12.5 K for each deposition temperature configuration to avoid thermal desorption of CO ice, and the interval between each cycle is controlled to be 3 minutes.

\section{\label{sec:res}Results and Discussion}
Since the dissociation energy of CO is 11.09 eV \citep{stamatovic70}, which is higher than the maximum photon energy of 10.88 eV provided by the MDHL, in these experiments, the triple bond of the CO molecule is only dissociated via the reaction
\begin{equation}
\rm CO+CO^*\to CO_2+C
\label{eq:toCO2}
\end{equation}
where CO$^*$ refers to a photoexcited molecule \citep{okabe78, gerakines1996ultraviolet, loeffler2005co2}. This process has low efficiency in our experiments, and the column density of CO$_2$ is always less than 5\% of the initial CO ice column density value, in agreement with previous works \citep{G10, chen14}.
The integrated ion current of desorbed CO molecules detected in the gas phase (m/z=28) during each CO ice irradiation cycle as a function of fluence shows the evolution of photodesorption, which is consistent with the decrease of the CO ice IR band.  
\subsection{Thickness effect} \label{subsec:thick}

\begin{figure*}
\centering
\includegraphics[width=18cm]{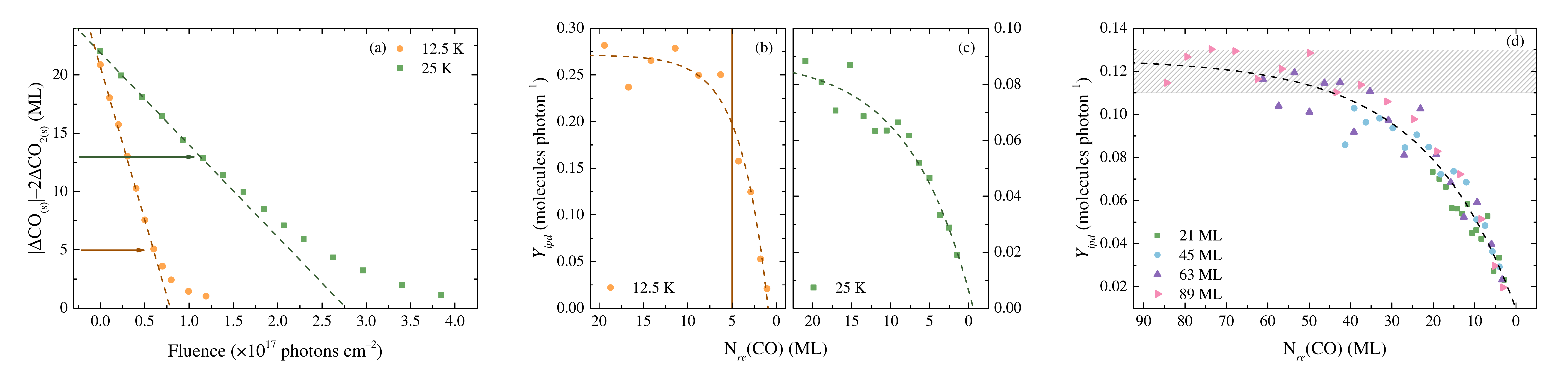}
\caption{(a) The photodesorption yield of 21 ML-thick CO ice calculated by carbon balance method as a function of fluence in the 12.5 and 25 K experiments. The dashed lines depict the linear photodepletion of CO ice resulting mainly from CO desorption. The arrows indicate the first data point that falls out of the linear fit. (b) and (c) $Y_{ipd}$ as a function of N$_{re}$(CO) in the 12.5 and 25 K experiments, respectively. (d) $Y_{ipd}$ as a function of N$_{re}$(CO) of four different ice thicknesses deposited at 25 K. The shaded region represents the saturated photodesorption yield.}
\label{F1}
\end{figure*}
The photodesorption yield of 21 ML-thick (1 ML $=10^{15}$ molecules cm$^{-2}$) CO ice is calculated with the carbon balance method $\rm |\Delta CO_{(s)}|-2\Delta CO_{2_{(s)}}$, since 2 CO molecules are needed to form a CO$_2$ molecule in solid phase, and the remaining fraction corresponds to the CO molecules photodesorbing from CO ice into gas phase. At 12.5 K deposition temperature, the photodesorption decreases linearly as a function of fluence during VUV-irradiation until the CO thickness is less than 5 ML, as shown in Fig. \ref{F1}(a). The turning point at 5 ML is consistent with \cite{G10} and \cite{chen14} with initial ice thickness less than 21 ML, indicating that the photodesorption process is no longer efficient since the thickness is not enough. We found an interesting result showing that the photodepletion of CO ice started to decrease already at a larger CO thickness in the 25 K experiment, since the linear drop fails when N$_{re}$(CO) is $\sim$12 ML. This effect of deposition temperature on the value of the ice thickness that corresponds to the turning point from a constant to a decreasing CO depletion rate has not been reported before. 
The general fitting method for deriving photodesorption yield will eliminate the variation of photodesorption in each irradiation period. A novel approach is proposed to explore the behavior of photodesorption called "instantaneous photodesorption yield" ($Y_{ipd}$), which is derived from the decrease of CO ice column density divided by the fluence in each single irradiation period to reveal the variation between each cycle. $Y_{ipd}$ is represented as a function of N$_{re}$(CO) in the 12.5 and 25 K experiments for each irradiation cycle, as in Fig. \ref{F1}(b) and (c). $Y_{ipd}$ for the 12.5 K deposition is rather constant, it falls within 0.24--0.28 molecules photon$^{-1}$, in the region that coincides with the linear fit in Fig. \ref{F1}(a) for CO ice column density above 5 ML (vertical line in Fig. \ref{F1}(b)). But only a gradually decreasing $Y_{ipd}$ is observed in the 25 K experiment. The initial CO ice thickness of 21 ML is therefore not sufficient to reach a constant photodesorption yield when CO is deposited at 25 K. Therefore, we experimented with four different thicknesses of CO ice: 21, 45, 63, and 89 ML were deposited at 25 K to reveal the thickness dependence on the photodesorption yield.

From 21 to 89 ML, $Y_{ipd}$ as a function of N$_{re}$(CO) follows an identical trend. Constant values within 0.11--0.13 molecules photon$^{-1}$ were measured if N$_{re}$(CO) was larger than 60 ML (Fig. \ref{F1}(d)). We therefore selected an ice thickness of 89 ML to continue our study of temperature dependence on the photodesorption of CO ice as follows.

\subsection{\label{subsec:temp}Temperature effect}
\begin{figure*}
\centering
\includegraphics[width=18cm]{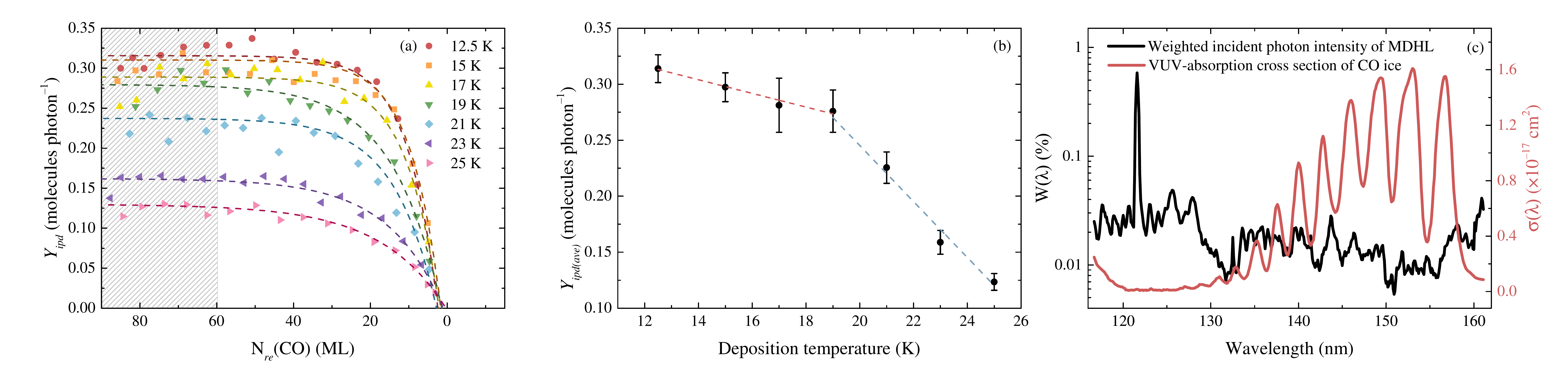}
\caption{(a) $Y_{ipd}$ as a function of N$_{re}$(CO) for CO deposited in the 12.5--25 K range. The shaded region represents the saturated photodesorption yield, becoming $Y_{ipd(ave)}$. (b) $Y_{ipd(ave)}$ as a function of deposition temperature for a 89 ML-thick CO ice. (c) $W(\lambda)$ and $\sigma(\lambda)$ of CO ice as a function of wavelength.}
\label{F2}
\end{figure*}
CO ice samples of 89 ML were deposited within the temperature range from 12.5 to 25 K to study the temperature effect on photodesorption yield. $Y_{ipd}$ as a function of N$_{re}$(CO) is represented in Fig. \ref{F2}(a). At the beginning of VUV-irradiation, $Y_{ipd}$ is rather constant in each configuration. $Y_{ipd}$ drops dramatically at a certain N$_{re}$(CO), which is dependent on deposition temperature. The average instantaneous photodesorption yield ($Y_{ipd(ave)}$) in the constant region (N$_{re}$(CO) $\ge$ 60 ML) represents the photodesorption yields shown in Fig. \ref{F2}(b), showing two different linear relations as a function of deposition temperature, with the turning point at 19 K. Above this temperature, $Y_{ipd(ave)}$ decreases faster. 

\subsection{\label{subsec:mech}Photodesorption estimation}
\begin{figure}
\centering
\includegraphics[width=8.5cm]{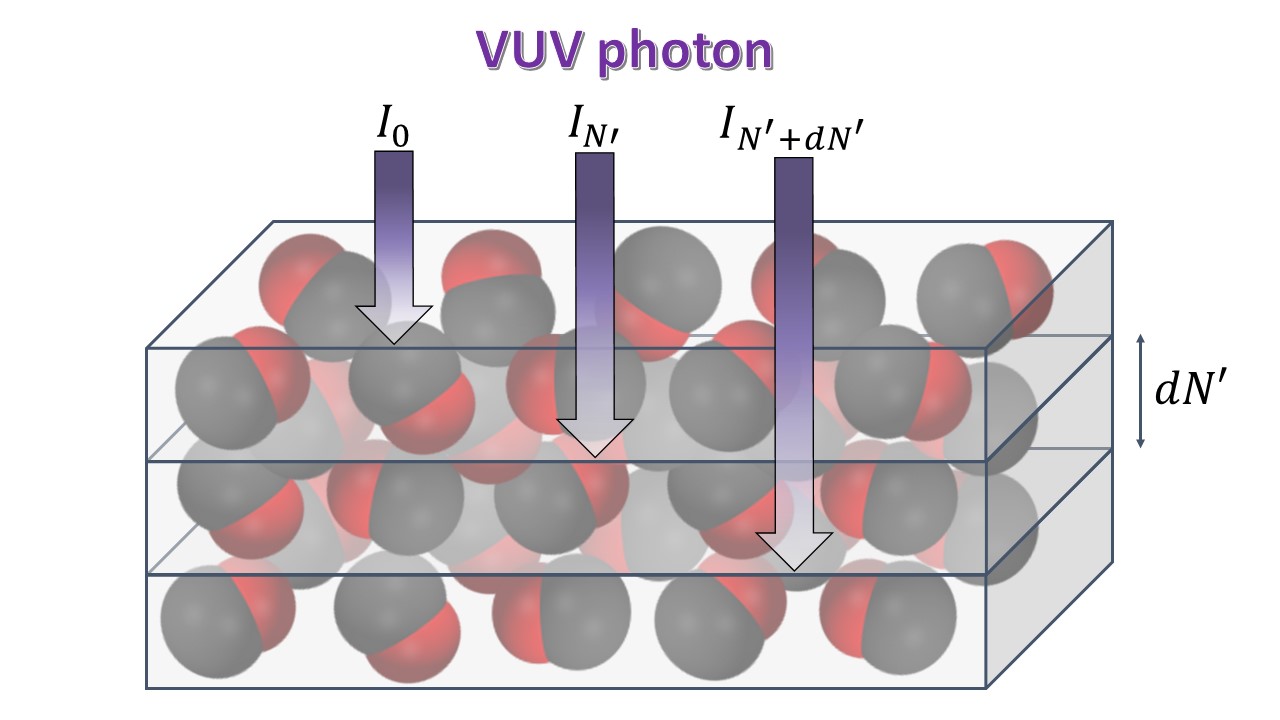}
\caption{
The schematic diagram of incident VUV photon passes through $N'$th layer. The subscript of $N$ represents the incident intensity passing through that layer, where 0 means the initial incident MDHL intensity.
}
\label{VUV}
\end{figure}
The first step in the photodesorption process is photon absorption in the CO ice, which is related to the photon energy distribution of the MDHL emission and its $\sigma(\lambda)$ at this range, depicted in Fig. \ref{F2}(c). The photodesorption yield is proportional to the summation of absorbed photons in each single $dN'$ layer, which can be calculated by the following equations. According to Beer--Lambert law, the transmitted intensity can be presented as 
\begin{equation} 
I_{N'}(\lambda, N')=W(\lambda)e^{-\sigma(\lambda)N'}
\label{eq:IN}
\end{equation}
where ${W(\lambda)=\frac{I_0(\lambda)}{\int I_0(\lambda) d\lambda}}$, the normalized incident photon intensity of MDHL, which means the integral over wavelength is normalized to unity. This allows us to know the ratio for each wavelength contributing to the photon intensity spectrum, $I_0(\lambda)$ the incident photon intensity, $\sigma(\lambda)$ the VUV-absorption cross-section of CO ice as a function of wavelength $\lambda$ in cm$^2$, $N'$ the column density of $N'$th layer in ML. $I_{N'}(\lambda, N')$ is the normalized transmittance intensity when the photons go through $N'$th layer as a function of wavelength, which can also be seen as the new incident intensity for the next $N'+dN'$ layer in Fig. \ref{VUV}. Since this photon intensity is dependent on the layer $N'$, the derivative with respect to $N'$th layer represents the wavelength weighted absorption ratio of $N'$th layer ($I_{abs}(\lambda,N')$), which
is still as a function of wavelength. Subsequently, the absorption ratio of $N'$th layer is integrated over wavelength which is the same as MDHL spectra to remove the relation with wavelength:
\begin{equation}
{I_{abs}(N')
=\int W(\lambda)\sigma(\lambda)e^{-\sigma(\lambda)N'}d\lambda}
\label{eq:intdIN}
\end{equation}

After the absorption of photon energy in the ice, molecules can desorb into the gas phase following two types of photodesorption: 1) photochemidesorption and 2) desorption induced by electronic transition (DIET) (e.g., \citealt{martin15}). The former could lead to the desorption of CO$_2$ molecules formed on the ice surface by the reaction in Eq.~\ref{eq:toCO2}, but the desorption of CO$_2$ molecules was not observed in our experiments, probably due to the low abundances, which is lower than the detection limit of the QMS used in these experiments. The direct dissociation of CO on the surface could lead to desorption of C and O atoms, but as we already mentioned this process requires a photon energy higher than the MgF$_2$ window cutoff frequency \citep{stamatovic70}. However, the C and O can be produced via Eq.~\ref{eq:toCO2} and dissociation of produced CO$_2$ in VUV irradiated CO ice, respectively \citep{chen14}.

But the method responsible for this energy transfer between neighboring CO molecules is not well understood. A theoretical study dedicated to CO photodesorption proposed various pathways, but only considered energy transfer between two close neighbors \citep{van2015molecular}. Experimentally, we showed that for CO ice deposited near 12.5 K the turning point from a constant to a decreasing $Y_{ipd}$ occurs at 5 ML ice column density, or about 12 ML in CO ice deposited at 25 K, suggesting that this is the average transmission distance, see Fig. \ref{F1}(a). Therefore, we know now that such a model must account for photon energy transfer between several molecules before the energy is fully dissipated, as it is expected for exciton propagation in the ice \citep{chen17}.

We propose here a method to describe the photodesorption of CO ice, including 3 parameters: $Y(T)$, $R_A(T)$, and $L(T)$. $Y_{ipd}$ can be seen as the accumulated $Y(T)$ from each single layer by considering the wavelength weighted absorption ratio as a function of $N'$th layer ($I_{abs}(N')$), and the energy transfer is assumed to be an exponentially decay-like behavior, written as $e^{\frac{-N'}{L(T)}}$. Moreover, $Y_{ipd}$ is also related to the flatness, and here we use relative effective surface area ($R_A(T)$) for analogy, where "relative" represents the concept of ratio since the effective surface area contains a unit, but the real value of this area is not available. $R_A(T)$ is on behalf of the surface flatness, when the surface is rugged, this value is larger to support more molecules desorbing out. $Y_{ipd}$ as a function of N$_{re}$(CO), shown in Fig. \ref{F2}(a), can be fitted with the equation including parameters adopted above: 
\begin{equation}
{Y_{ipd}(N,T)=\int_{0}^{N}I_{abs}(N')\times e^{\frac{-N'}{L(T)}}\times Y(T)\times R_A(T)~dN'
\label{eq:Y}}
\end{equation}
where the first term is wavelength weighted absorption ratio of $N'$th layer in ML$^{-1}$ as introduced before, and $L(T)$ the energy transfer length in ML, $N$ the column density of CO ice in ML, $Y(T)$ the single ice layer contributed desorption yield in molecules photon$^{-1}$, which means the photodesorption yield contributed by a single layer after absorbing photons. $Y(T)$, $L(T)$, and $R_A(T)$ are the parameters dependent on deposition temperature. Except $I_{abs}(N')$, which can be calculated from Eq.~\ref{eq:IN} to \ref{eq:intdIN}, the other three parameters should be fitted to obtain the values shown in Fig. \ref{Fit}.

\subsection{Dominating parameters for photodesorption}
\begin{figure}
\centering
\includegraphics[width=8.5cm]{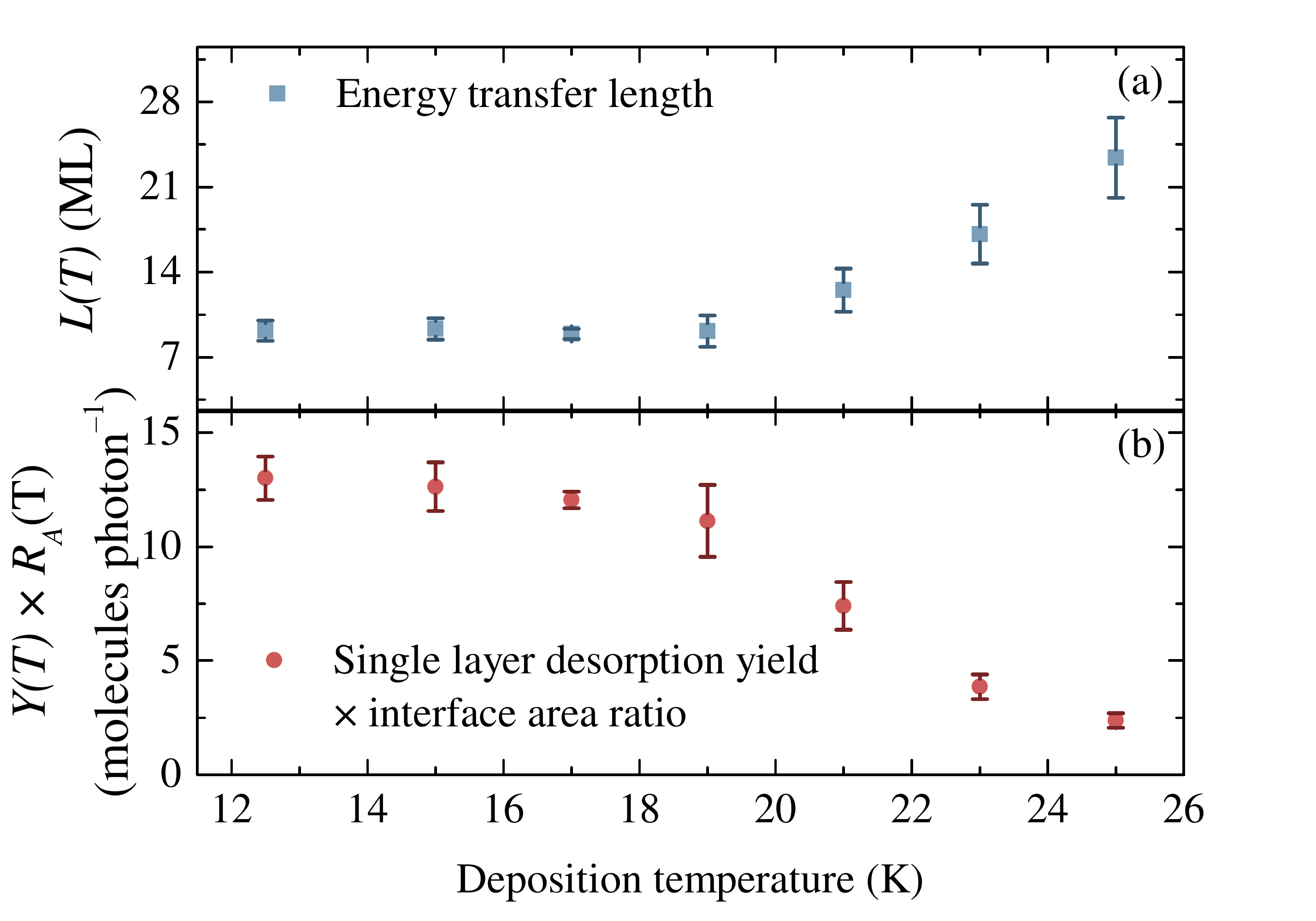}
\caption{(a) $L(T)$ and (b) $Y(T)$ times $R_A(T)$ as a function of deposition temperature.}
\label{Fit}
\end{figure}
$L(T)$ represents the energy absorbed along distance L, there is only 37\% ($e^{-1}$) of initial absorbed energy left since the energy dissipates exponentially along the penetration, which is related to the structural properties of CO ice. At 12.5 K deposition temperature, CO ice presents an amorphous structure. As deposition temperature increases, the structure becomes partially crystalline with a phase transition at 20 K \citep{kouchi90,kouchi2020direct}. Below phase transition temperature, the structure of CO ice is the same, leading to a constant $L(T)$. The CO molecules are arranged in a nearly crystalline structure at higher deposition temperature, i.e. above 20 K, $L(T)$ is thus larger, as shown in Fig. \ref{Fit}(a). In addition, \cite{kouchi2021transmission} shows that the crystalline $\alpha$-CO is formed above 21 K, supported by the TEM images and the electron diffraction pattern, and the structure of CO ice does not only depend on the temperature because also the substrate used for CO ice deposition plays a role: the size of the three-dimensional CO island can reach $\approx$ 200 nm when deposited on amorphous H$_2$O ice, while it is only $\approx$ 50 nm on CO$_2$ ice. The substrate is CaF$_2$ for all deposition temperature configurations in this study. The substrate effect on the crystallization of CO ice is out of the scope of this paper The lower energy transfer length below 20K is related to the non-alignment of molecular dipole, hindering resonant transfer mechanisms based on dipole-dipole interactions. In addition, this parameter determines the photodesorption trend as a function of N$_{re}$(CO), which means that the photodesorption yield drops either dramatically or gradually at low and high deposition temperatures, respectively. 

Due to the mathematical limitations, the parameters $Y(T)$ and $R_A(T)$ cannot be estimated separately, but the product $Y(T) \times R_A(T)$ can be obtained. Both $Y(T)$ and $R_A(T)$ are related to the CO ice growth temperature. Therefore, the formation temperature of CO ice dominates $Y(T) \times R_A(T)$. This product of $Y(T)$ and $R_A(T)$ remained almost constant below 19 K deposition temperatures and decreased linearly above 19 K. The declining trend from 19 to 25 K is referred to the $E_S$ in CO ice, which is reported to decrease linearly in this deposition temperature range between 20--26 K \citep{lasne15}. An observable effect of this $E_S$ is the peak red-shift in VUV-absorption spectra of solid CO starting above 20 K \citep{chen17}. The orientation and molecular disorder of CO ice deposited at different temperatures are also discussed in \cite{carrascosa2021intriguing}. Indeed, the excited states of CO molecules in the ice are shifted to a lower level as a result of lower $E_S$ at deposition temperatures above 20 K, leading to a decreasing single ice layer contributed desorption yield. On the other hand, for the nonpolar CO$_2$ ice, the photodesorption yield is independent on the deposition temperature \citep{sie19}, and the VUV absorption spectra also show no peak shift, which indicates that no spontelectric field is generated in CO$_2$ ice. Also no peak shift of the VUV absorption bands was reported for another nonpolar molecule, N$_2$; deposited at 14, 20, and 24 K also shows a same result of VUV absorption spectra \citep{chen17}. Below 19 K, the $E_S$ is constant, therefore the $R_A(T)$ is the primary parameter influencing the photodesorption yield at the range of deposition temperature of 12.5--19 K. When the surface is flat, this value is assumed to be 1, while it becomes larger as the effective surface area is larger. When CO ice is deposited at low temperature, the porosity is larger compared to that at high temperature, leading to a larger $R_A(T)$, causing the higher photodesorption yield value in Fig. \ref{F2}(b). This is supported by \cite{cazaux2017}, the binding energy of CO ice deposited at 14 K (Fig. 14) and 6 K (Fig. 15) are calculated by Monte Carlo simulations and the results show that gas phase CO molecules can stick where they arrive without having to rearrange at very low temperatures. This leads to a large distribution of binding energy and many pores can be observed in such ices, when the number of pores is higher, the effective surface area is thought to be also higher. 
\section{\label{sec:astro}Astrophysical Implication and Conclusions}
There is substantial literature dedicated to exploring the photodesorption of solid CO ice in the laboratory, and the yields span several scales from $10^{-3}$ to $10^{-1}$ molecules photon$^{-1}$ among these works \citep{oberg09, fayolle11, chen14, G16, paardekooper16}, with different CO ice thicknesses, deposition temperatures, and irradiation configurations. As in \cite{chen14} and \cite{sie19}, the incident energy distribution of MDHL plays an important role in discussing the absorption of CO ice, which would lead to distinct photodesorption behavior.

In the present paper, the novel calculation of $Y_{ipd}$ is introduced for the ice thickness and deposition temperature dependence on the photodesorption yield of CO ice, which is discussed along with the method of photodesorption at different deposition temperatures dominated by three parameters: energy transfer length, single ice layer contributed desorption yield, and relative effective surface area. The ice thickness taking part in the photodesorption process to support a constant photodesorption yield of CO ice is deposition temperature dependent. At higher deposition temperatures, a thicker ice is required to reach a maximum and constant value of the photodesorption yield. At 25 K, the instantaneous photodesorption yield becomes constant when the thickness of CO ice is larger than 60 ML. Instead, when the ice thickness is less than 30 ML, the instantaneous photodesorption yield drops drastically. 

In the current model of astrophysical ice mantles covering dust grains, such as those observed toward dense interstellar clouds and cold circumstellar regions, the outer layer is made of the most volatile molecules, being CO the most abundant species, which likely coexists with N$_2$ and H$_2$ that have very low UV-absorption cross-sections \citep{martin2020exploring}. CO molecules are also observed in the gas phase of the coldest regions in dense clouds, suggesting that a strong accretion is counteracted by non-thermal desorption processes. The reported experimental yields can be applied to the CO photodesorption processes induced by the secondary UV-field in the interior of dense interstellar clouds. While chemical desorption would operate in bare dust grain surfaces \citep{minissale2016dust}, desorption of molecules in ice-covered grains is likely driven by cosmic rays and energetic (UV) photons. We found that the photodesorption yield of CO ice is strongly temperature and ice thickness dependent. X rays may dominate the ice processing in young protoplanetary disks \citep{ciaravella2020x,jimenez2022x}, this could restructure the ice mantles and change the $R_A(T)$ parameter. N(CO) used as input in the model of a dense cloud interior is related to the cloud lifetime, but this value reaches a maximum for lifetimes shorter than the total lifetime of the cloud \citep{G10}. A typical ice mantle thickness of 10 nm or 30 ML, of which only the uppermost layers are composed of CO, will show a reduced photodesorption yield, since we report that the maximum photodesorption occurs for ice thickness above 60 ML. This ice thickness might be attained in precometary icy grains. On the other hand, the low growth temperature near 10 K of interstellar ice mantles will favor the photodesorption, while the temperature gradient in disks will lead to a varying photodesorption of the ice mantles covering a broader range of growth temperatures and ice thickness distribution. Current astrochemical disk models are complex as they take into account numerous parameters (e.g., \citealt{walsh2010chemical,woitke2018modelling}). These models can include our experimental findings as input to improve their treatment of CO photodesorption. \cite{woitke2018modelling} adapted their model in the past based on our previous results and \cite{cazaux2017, cazaux2022photoprocessing} used Monte Carlo to interpret our experimental results and extend them to the astrophysical scenario. The extrapolation of our experimental model to an astrophysical one that simulates a disk environment or a dense interstellar cloud interior, where ice mantles are exposed to UV photons, is beyond the scope of this paper and may be performed in collaboration with modellers.

\begin{acknowledgments}
This work benefited from financial support from grant No. NSTC 110-2628-008-004-MY4 (Y.-J.C.). The Spanish Ministry of Science, Innovation and Universities supported this research under Grant PID2020-118974GB-C21 (AEI/FEDER, EU), Program MDM-2017-0737 Unidad de Excelencia “María de Maeztu”–Centro de Astrobiología (INTA-CSIC).
\end{acknowledgments}

\vspace{5mm}

\bibliography{sample631}{}

\begin{thebibliography}{}
\expandafter\ifx\csname natexlab\endcsname\relax\def\natexlab#1{#1}\fi
\providecommand{\url}[1]{\href{#1}{#1}}
\providecommand{\dodoi}[1]{doi:~\href{http://doi.org/#1}{\nolinkurl{#1}}}
\providecommand{\doeprint}[1]{\href{http://ascl.net/#1}{\nolinkurl{http://ascl.net/#1}}}
\providecommand{\doarXiv}[1]{\href{https://arxiv.org/abs/#1}{\nolinkurl{https://arxiv.org/abs/#1}}}

\bibitem[{Avouris \& Walkup(1989)}]{avouris1989fundamental}
Avouris, P., \& Walkup, R.~E. 1989, ARPC, 40, 173

\bibitem[{Bertin {et~al.}(2012)Bertin, Fayolle, Romanzin, {\"O}berg, Michaut,
  Moudens, Philippe, Jeseck, Linnartz, \& Fillion}]{bertin2012uv}
Bertin, M., Fayolle, E.~C., Romanzin, C., {et~al.} 2012, PCCP, 14, 9929

\bibitem[{Bonn {et~al.}(1999)Bonn, Funk, Hess, Denzler, Stampfl, Scheffler,
  Wolf, \& Ertl}]{bonn1999phonon}
Bonn, M., Funk, S., Hess, C., {et~al.} 1999, Sci, 285, 1042

\bibitem[{Boogert {et~al.}(2015)Boogert, Gerakines, \&
  Whittet}]{boogert2015observations}
Boogert, A.~A., Gerakines, P.~A., \& Whittet, D.~C. 2015, Annual Review of
  Astronomy and Astrophysics, 53, 541

\bibitem[{Carrascosa {et~al.}(2021)Carrascosa, Caro, Gonz{\'a}lez-d{\'\i}az,
  Suevos, \& Chen}]{carrascosa2021intriguing}
Carrascosa, H., Caro, G.~M., Gonz{\'a}lez-d{\'\i}az, C., Suevos, J., \& Chen,
  Y.-J. 2021, \apj, 916, 1

\bibitem[{Cazaux {et~al.}(2022)Cazaux, Carrascosa, Mu{\~n}oz~Caro, Caselli,
  Fuente, Navarro-Almaida, \&
  Rivi{\'e}re-Marichalar}]{cazaux2022photoprocessing}
Cazaux, S., Carrascosa, H., Mu{\~n}oz~Caro, G., {et~al.} 2022, \aap, 657, A100

\bibitem[{Cazaux {et~al.}(2017)Cazaux, Mart{\'\i}n-Dom{\'e}nech, Chen,
  Mu\~{n}oz Caro, \& D{\'\i}az}]{cazaux2017}
Cazaux, S., Mart{\'\i}n-Dom{\'e}nech, R., Chen, Y., Mu\~{n}oz Caro, G.~M., \&
  D{\'\i}az, C.~G. 2017, \apj, 849, 80

\bibitem[{Cecchi-Pestellini \& Aiello(1992)}]{cecchi1992cosmic}
Cecchi-Pestellini, C., \& Aiello, S. 1992, \mnras, 258, 125

\bibitem[{Chen {et~al.}(2014)Chen, Chuang, Mu\~{n}oz Caro, Nuevo, Chu, Yih, Ip,
  \& Wu}]{chen14}
Chen, Y.-J., Chuang, K.-J., Mu\~{n}oz Caro, G.~M., {et~al.} 2014, \apj, 781, 15

\bibitem[{Chen {et~al.}(2017)Chen, Mu\~{n}oz Caro, Aparicio,
  Jim{\'e}nez-Escobar, Lasne, Rosu-Finsen, McCoustra, Cassidy, \&
  Field}]{chen17}
Chen, Y.-J., Mu\~{n}oz Caro, G.~M., Aparicio, S., {et~al.} 2017, \prl, 119,
  157703

\bibitem[{Ciaravella {et~al.}(2020)Ciaravella, Caro, Jim{\'e}nez-Escobar,
  Cecchi-Pestellini, Hsiao, Huang, \& Chen}]{ciaravella2020x}
Ciaravella, A., Caro, G. M.~M., Jim{\'e}nez-Escobar, A., {et~al.} 2020, PNAS

\bibitem[{Cruz-Diaz {et~al.}(2014)Cruz-Diaz, Mu\~{n}oz Caro, Chen, \&
  Yih}]{cruz14p}
Cruz-Diaz, G., Mu\~{n}oz Caro, G.~M., Chen, Y.-J., \& Yih, T.-S. 2014, \aap,
  562, A119

\bibitem[{Dartois {et~al.}(2018)Dartois, Chabot, Barkach, Rothard, Aug{\'e},
  Agnihotri, Domaracka, \& Boduch}]{dartois2018cosmic}
Dartois, E., Chabot, M., Barkach, T.~I., {et~al.} 2018, \aap, 618, A173

\bibitem[{Fayolle {et~al.}(2011)Fayolle, Bertin, Romanzin, Michaut, {\"O}berg,
  Linnartz, \& Fillion}]{fayolle11}
Fayolle, E.~C., Bertin, M., Romanzin, C., {et~al.} 2011, \apjl, 739, L36

\bibitem[{Funk {et~al.}(2000)Funk, Bonn, Denzler, Hess, Wolf, \&
  Ertl}]{funk2000desorption}
Funk, S., Bonn, M., Denzler, D.~N., {et~al.} 2000, \jcp, 112, 9888

\bibitem[{Gerakines {et~al.}(1996)Gerakines, Schutte, \&
  Ehrenfreund}]{gerakines1996ultraviolet}
Gerakines, P., Schutte, W., \& Ehrenfreund, P. 1996, \aap, 312, 289

\bibitem[{Gonz{\'a}lez~D{\'\i}az {et~al.}(2019)Gonz{\'a}lez~D{\'\i}az,
  Carrascosa~de Lucas, Aparicio, Mu\~{n}oz Caro, Sie, Hsiao, Cazaux, \&
  Chen}]{Cristobal19}
Gonz{\'a}lez~D{\'\i}az, C., Carrascosa~de Lucas, H., Aparicio, S., {et~al.}
  2019, \mnras, 486, 5519

\bibitem[{Gredel {et~al.}(1989)Gredel, Lepp, Dalgarno, \& Herbst}]{gredel1989}
Gredel, R., Lepp, S., Dalgarno, A., \& Herbst, E. 1989, \apj, 347, 289

\bibitem[{Jiang {et~al.}(1975)Jiang, Person, \& Brown}]{jiang1975}
Jiang, G.~J., Person, W.~B., \& Brown, K.~G. 1975, \jcp, 62, 1201

\bibitem[{Jim{\'e}nez-Escobar {et~al.}(2022)Jim{\'e}nez-Escobar, Ciaravella,
  Cecchi-Pestellini, Caro, Huang, Sie, \& Chen}]{jimenez2022x}
Jim{\'e}nez-Escobar, A., Ciaravella, A., Cecchi-Pestellini, C., {et~al.} 2022,
  \apj, 926, 176

\bibitem[{Kouchi(1990)}]{kouchi90}
Kouchi, A. 1990, JCrGr, 99, 1220

\bibitem[{Kouchi {et~al.}(2020)Kouchi, Furuya, Hama, Chigai, Kozasa, \&
  Watanabe}]{kouchi2020direct}
Kouchi, A., Furuya, K., Hama, T., {et~al.} 2020, \apjl, 891, L22

\bibitem[{Kouchi {et~al.}(2021)Kouchi, Tsuge, Hama, Oba, Okuzumi, Sirono,
  Momose, Nakatani, Furuya, Shimonishi, {et~al.}}]{kouchi2021transmission}
Kouchi, A., Tsuge, M., Hama, T., {et~al.} 2021, \apj, 918, 45

\bibitem[{Lasne {et~al.}(2015)Lasne, Rosu-Finsen, Cassidy, McCoustra, \&
  Field}]{lasne15}
Lasne, J., Rosu-Finsen, A., Cassidy, A., McCoustra, M.~R., \& Field, D. 2015,
  PCCP, 17, 30177

\bibitem[{Loeffler {et~al.}(2005)Loeffler, Baratta, Palumbo, Strazzulla, \&
  Baragiola}]{loeffler2005co2}
Loeffler, M., Baratta, G., Palumbo, M., Strazzulla, G., \& Baragiola, R. 2005,
  \aap, 435, 587

\bibitem[{Madey(1994)}]{madey1994}
Madey, T.~E. 1994, SurSc, 299, 824

\bibitem[{Mart{\'\i}n-Dom{\'e}nech {et~al.}(2020)Mart{\'\i}n-Dom{\'e}nech,
  Maksiutenko, {\"O}berg, \& Rajappan}]{martin2020exploring}
Mart{\'\i}n-Dom{\'e}nech, R., Maksiutenko, P., {\"O}berg, K.~I., \& Rajappan,
  M. 2020, \apj, 902, 116

\bibitem[{Mart{\'\i}n-Dom{\'e}nech {et~al.}(2015)Mart{\'\i}n-Dom{\'e}nech,
  Manzano-Santamar{\'\i}a, Mu\~{n}oz Caro, Cruz-D{\'\i}az, Chen, Herrero, \&
  Tanarro}]{martin15}
Mart{\'\i}n-Dom{\'e}nech, R., Manzano-Santamar{\'\i}a, J., Mu\~{n}oz Caro,
  G.~M., {et~al.} 2015, \aap, 584, A14

\bibitem[{Minissale {et~al.}(2016)Minissale, Dulieu, Cazaux, \&
  Hocuk}]{minissale2016dust}
Minissale, M., Dulieu, F., Cazaux, S., \& Hocuk, S. 2016, \aap, 585, A24

\bibitem[{Mu\~{n}oz Caro {et~al.}(2016)Mu\~{n}oz Caro, Chen, Aparicio,
  Jim{\'e}nez-Escobar, Rosu-Finsen, Lasne, \& McCoustra}]{G16}
Mu\~{n}oz Caro, G.~M., Chen, Y.-J., Aparicio, S., {et~al.} 2016, \aap, 589, A19

\bibitem[{Mu\~{n}oz Caro {et~al.}(2010)Mu\~{n}oz Caro, Jim{\'e}nez-Escobar,
  Mart{\'\i}n-Gago, Rogero, Atienza, Puertas, Sobrado, \& Torres-Redondo}]{G10}
Mu\~{n}oz Caro, G.~M., Jim{\'e}nez-Escobar, A., Mart{\'\i}n-Gago, J., {et~al.}
  2010, \aap, 522, A108

\bibitem[{{\"O}berg {et~al.}(2009){\"O}berg, Van~Dishoeck, \&
  Linnartz}]{oberg09}
{\"O}berg, K.~I., Van~Dishoeck, E.~F., \& Linnartz, H. 2009, \aap, 496, 281

\bibitem[{Okabe {et~al.}(1978)}]{okabe78}
Okabe, H., {et~al.} 1978, Photochemistry of small molecules, Vol. 431 (Wiley
  New York)

\bibitem[{Paardekooper {et~al.}(2016)Paardekooper, Bossa, \&
  Linnartz}]{paardekooper16}
Paardekooper, D., Bossa, J.-B., \& Linnartz, H. 2016, \aap, 592, A67

\bibitem[{Penteado {et~al.}(2015)Penteado, Boogert, Pontoppidan, Ioppolo,
  Blake, \& Cuppen}]{penteado2015spectroscopic}
Penteado, E., Boogert, A., Pontoppidan, K., {et~al.} 2015, \mnras, 454, 531

\bibitem[{Pontoppidan {et~al.}(2004)Pontoppidan, Van~Dishoeck, \&
  Dartois}]{pontoppidan2004mapping}
Pontoppidan, K., Van~Dishoeck, E., \& Dartois, E. 2004, \aap, 426, 925

\bibitem[{Pontoppidan {et~al.}(2003)Pontoppidan, Fraser, Dartois, Thi,
  Van~Dishoeck, Boogert, d'Hendecourt, Tielens, \& Bisschop}]{pontoppidan2003m}
Pontoppidan, K., Fraser, H., Dartois, E., {et~al.} 2003, \aap, 408, 981

\bibitem[{Reimann {et~al.}(1988)Reimann, Brown, \&
  Johnson}]{reimann1988electronically}
Reimann, C., Brown, W., \& Johnson, R. 1988, \prb, 37, 1455

\bibitem[{Schwoerer \& Wolf(2007)}]{schwoerer2007organic}
Schwoerer, M., \& Wolf, H.~C. 2007, Organic molecular solids (John Wiley \&
  Sons)

\bibitem[{Shen {et~al.}(2004)Shen, Greenberg, Schutte, \&
  Van~Dishoeck}]{shen2004cosmic}
Shen, C., Greenberg, J., Schutte, W., \& Van~Dishoeck, E. 2004, \aap, 415, 203

\bibitem[{Sie {et~al.}(2019)Sie, Mu\~{n}oz Caro, Huang,
  Mart{\'\i}n-Dom{\'e}nech, Fuente, \& Chen}]{sie19}
Sie, N.-E., Mu\~{n}oz Caro, G.~M., Huang, Z.-H., {et~al.} 2019, \apj, 874, 35

\bibitem[{Stamatovic \& Schulz(1970)}]{stamatovic70}
Stamatovic, A., \& Schulz, G. 1970, \jcp, 53, 2663

\bibitem[{van Hemert {et~al.}(2015)van Hemert, Takahashi, \& van
  Dishoeck}]{van2015molecular}
van Hemert, M.~C., Takahashi, J., \& van Dishoeck, E.~F. 2015, JPCA, 119, 6354

\bibitem[{Walsh {et~al.}(2010)Walsh, Millar, \& Nomura}]{walsh2010chemical}
Walsh, C., Millar, T., \& Nomura, H. 2010, \apj, 722, 1607

\bibitem[{Woitke {et~al.}(2018)Woitke, Min, Thi, Roberts, Carmona, Kamp,
  M{\'e}nard, \& Pinte}]{woitke2018modelling}
Woitke, P., Min, M., Thi, W.-F., {et~al.} 2018, \aap, 618, A57

\end{thebibliography}
\bibliographystyle{aasjournal}



\end{document}